\documentclass[aps,pre,twocolumn,superscriptaddress,showpacs]{revtex4-1}

\usepackage{amsmath}
\usepackage{amssymb}
\usepackage{hyperref}
\usepackage[arrow, matrix, curve]{xy}
\usepackage{bm}
\usepackage{yhmath}
\usepackage{color}

\DeclareMathOperator\Real{Re} 
\DeclareMathOperator\Imag{Im}
\DeclareMathOperator*\limav{lim\,av}

\usepackage{color}
\newcommand\diff{\mathrm{d}}

\newtheorem{theorem}{Theorem}[section]
\newtheorem{lemma}[theorem]{Lemma}

\usepackage[usenames,dvipsnames]{xcolor}
\hypersetup{colorlinks=true, linkcolor=BrickRed, urlcolor=blue!50!black, citecolor=blue!50!black}

\usepackage[normalem]{ulem}
\makeatletter
\newcommand\hide@visible[1]{%
  \bgroup\fboxsep=.3ex\colorbox{Gray}{begin hide}%
  #1\colorbox{Gray}{end hide}\egroup%
}
\newcommand\hide@hidden[1]{%
  \bgroup\fboxsep=.3ex\colorbox{Gray}{hidden text}%
}
\newcommand\hide@invisible[1]{}
\newcommand\makevisible{\let\hide\hide@visible}
\newcommand\makehidden{\let\hide\hide@hidden}
\newcommand\makeinvisible{\let\hide\hide@invisible}
\makeatother
\makehidden

\begin{document}

% Use the \preprint command to place your local institutional report
% number in the upper righthand corner of the title page in preprint mode.
% Multiple \preprint commands are allowed.
% Use the 'preprintnumbers' class option to override journal defaults
% to display numbers if necessary
%\preprint{}

%Title of paper
\title{Long-time limit of correlation functions}

% \affiliation can be followed by \email, \homepage, \thanks as well.

\author{Thomas Franosch}
\affiliation{Institut f\"ur Theoretische Physik, Leopold-Franzens-Universit\"at Innsbruck, Technikerstra{\ss}e~25/2,  A-6020 Innsbruck, Austria}
\email[]{thomas.franosch@uibk.ac.at}
%\homepage[]{Your web page}
%\thanks{}
%\altaffiliation{}

%Collaboration name if desired (requires use of superscriptaddress
%option in \documentclass). \noaffiliation is required (may also be
%used with the \author command).
%\collaboration can be followed by \email, \homepage, \thanks as well.
%\collaboration{}
%\noaffiliation
 
\date{\today}

% insert suggested PACS numbers in braces on next line
\pacs{02.50.Ey, 02.60.Nm, 61.20.Lc, 64.70.Q- }
%02.50.Ey 	Stochastic processes
%02.60.Nm 	Integral and integrodifferential equations
%61.20.Lc 	Time-dependent properties; relaxation (for glass transitions, see 64.70.P-)
%64.70.Q- 	Theory and modeling of the glass transition

% insert suggested keywords - APS authors don't need to do this
\keywords{Stochastic processes, time-dependent correlation functions, glass transition}

%\maketitle must follow title, authors, abstract, \pacs, and \keywords

\begin{abstract}
Auto-correlation functions in an equilibrium stochastic process are well-characterized by Bochner's theorem as Fourier transforms of a finite symmetric Borel measure. The existence of a long-time limit of these correlation functions depends on the spectral properties of the measure. Here we provide conditions applicable to a wide-class of dynamical theories  guaranteeing the existence of the long-time limit. 
We discuss the implications in the context of the mode-coupling theory of the glass transition where a non-trivial long-time limit signals an idealized glass state.
\end{abstract}

\maketitle

% body of paper here - Use proper section commands
% References should be done using the \cite, \ref, and \label commands

\section{Introduction}
Stochastic processes constitute the basis of virtually all dynamics and therefore are of  fundamental importance, in particular in condensed matter physics. There the quantities of interest are dynamic correlation functions, for example via the fundamental fluctuation-dissipation theorem~\cite{KuboBuch:1991}, they encode  the linear response with respect to an external perturbation. Furthermore, scattering techniques employing neutrons, light, or X-rays,   directly probe the dynamics via time-dependent correlation functions~\cite{Hansen:Theory_of_Simple_Liquids}. 

Physically one expects that all observables are coupled to their environment consisting of many degrees of freedom such that all dynamic processes are exposed to some damping mechanism. Then correlations should decay at large times and display only trivial behavior. In particular, persistent oscillations, quasi-periodic or intermittent behavior should not occur. The usual scenario is thus that correlation functions approach zero at large times which corresponds to an ergodic behavior. A fingerprint of non-ergodic behavior is thus if correlation functions display a non-trivial long-time limit. In the context of liquid dynamics~\cite{Goetze:Complex_Dynamics} %or disordered spin systems~\cite{?}
 this non-ergodic behavior is associated with an idealized glass state, where part of the initial correlation remains frozen-in for arbitrarily large times.  In a scattering experiment a non-trivial long-time limit yields an elastic line at zero frequencies broadened only by the instrument resolution.  

The existence of a long-time limit for correlation functions appears to be non-trivial, in particular, peculiar examples for correlation functions (in the mathematical literature refered to as covariance functions or  characteristic functions of a finite measure) are known that neither display a limit,  nor are they almost periodic (see ~\cite{Lukacs:Characteristic_Functions}, p. 27, Sec. 2.2).

A common strategy to develop a theory for time-correlation functions is to derive equations of motion encoding the dynamics of the underlying system. For example,
 the Zwanzig-Mori projection operator formalism yields formally exact equations of motion where the complexity arising due to the many-body interaction is deferred to a memory kernel~\cite{KuboBuch:1991,Hansen:Theory_of_Simple_Liquids,Goetze:Complex_Dynamics}. Then the question of the existence of a (non-trivial) long-time limit for the correlation function
 is a property inherited from the memory kernel. 

The goal of this work is to provide rather general conditions on equations of motion  such that the long-time limit of the correlation function is guaranteed to exist. Such equations naturally emerge for classical many-particle systems where the time evolution is driven by Newtonian dynamics. In particular, they have been the starting point for the mode-coupling theory of the glass transition~\cite{Goetze:Complex_Dynamics} which yields a series of testable non-trivial predictions of the dynamics as a liquid is driven towards a non-ergodic state. 

The paper is organized as follows, in Sec.~\ref{Sec:correlation_functions} basic properties of time correlation functions are recalled, The main contribution of this work is in Sec.~\ref{Sec:Model_equations} where the problem is formulated and the existence of a long-time limit is demonstrated. The connection to the mode-coupling theory of the glass transition is discussed in Sec.~\ref{Sec:MCT}. At the end, conclusions and possible generalizations are provided.
  
\section{Correlation functions in equilibrium}\label{Sec:correlation_functions}

For complex-valued observables $A(t)$, probability theory provides constraints on the class of admissable functions that correspond to (equilibrium) correlation functions $S(t) = \langle \delta A(t)^* \delta A(0) \rangle$~\cite{Feller:Probability_2} in a stationary stochastic process. Here $\delta A(t) = A(t)- \langle A(t) \rangle$ denotes the fluctuation around its stationary mean. In particular, employing time-translation invariance, one  verifies that correlation functions are positive definite in the following sense: Given a finite set of complex numbers $\xi_i$ and real times $t_i$ the quantity $\sum_{i,j} \xi_i^* S(t_i-t_j) \xi_j \geq 0$ is non-negative for any correlation function. By Bochner's theorem, the converse is also true and therefore provides a complete characterization of admissible functions. In equilibrium processes, which we define as stationary stochastic processes such that the time-reversed trajectory is equally likely, correlation functions are furthermore real and symmetric in time, $S(t) = S(t)^* = S(-t)$. By the spectral representation theorem~\cite{Feller:Probability_2}, correlation functions can then be written as characteristic functions of a  finite Borel measure $\sigma(\cdot)$ 
\begin{equation}
 S(t) = \int_\mathbb{R} e^{-i \omega t} \sigma(\diff \omega),
\end{equation}
and for equilibrium correlation functions the measure can be chosen to be symmetric. In particular, one infers that correlation functions are bounded by their initial value $|S(t)| \leq S(t=0)=\sigma(\mathbb{R})$ and are uniformly continuous.

It will be useful to  define the Fourier-Laplace transform by
\begin{equation}
\hat{S}(z) = \text{i} \int_0^\infty e^{\text{i} z t} S(t) \diff t,
\end{equation}
where the complex frequency $z$ is confined to the upper complex half plane $\mathbb{C}_+ := \{ z \in \mathbb{C} | \text{Im}[z] >0 \}$. The spectral representation theorem implies 
\begin{equation}\label{eq:Borel_transform}
 \hat{S}(z) =  \int \frac{1 }{\omega - z}  \sigma(\diff \omega),
\end{equation}
i.e. $\hat{S}(z)$ is  the Borel-Stieltjes transform of the finite symmetric measure $\sigma$.
The following properties follow readily   from the preceding relation for the Fourier-Laplace transforms of correlation functions in equilibrium
\begin{enumerate}
\item[(1)] $\hat{S}(z)$ is complex analytic for $z\in \mathbb{C}_+$
\item[(2)]  $\hat{S}( -z^*) = -\hat{S}(z)^*$ 
\item[(3)]  $\lim_{\eta \to \infty} \eta \Imag[\hat{S}(z = i\eta)]$ is finite 
\item[(4)]  $\Imag[ \hat{S}(z)]  \geq 0$ for $z\in \mathbb{C}_+$
\end{enumerate}
Conversely,  the Riesz-Herglotz representation theorem~\footnote{see N.I. Akhiezer, \emph{The classical moment problem} (Oliver \& Boyd, 1965) p. 92-93, Sec. 3.1., Eq. [3.3a] or F.~Gesztesy and E.~Tsekanovskii, \emph{Math. Nach.}, {\bf 218}, (2000), p.65, Sec. 2, Theorem 2.2}
%[see e.g. Akhiezer~\cite{Akhiezer:Classical_Moment_Problem}, p. 92-93, Sec. 3.1., Eq. [3.3a], Gesztesy~\cite{Gesztesy:2000}, p. 65, Sec. 2, Theorem 2.2] 
reveals that these four conditions  are sufficient for $\hat{S}(z)$ being the Laplace transform of an equilibrium correlation function. Functions with these properties are discussed in the literature under the names of  Nevanlinna, Herglotz, and Pick.

\section{Model equations and spectral properties}~\label{Sec:Model_equations}  

We consider equations of motion for the real function $S: t \in [0,\infty[ \mapsto S(t) \in \mathbb{R}$
\begin{equation}\label{eq:harmonic_oscillator}
J^{-1} \ddot{S}(t) + \nu  \dot{S}(t) +  S^{-1} S(t) + \int_0^t  M(t-t')  \dot{S}(t') \diff t' =0 ,
\end{equation}
with parameters $S> 0, \nu > 0, J >0 $ and  initial conditions $S(t=0) = S$, $\dot{S}(t=0) = 0$. 
Furthermore, the kernel $M(t)$ is assumed to be an equilibrium correlation function, i.e. by the spectral representation theorem~\cite{Feller:Probability_2} 
\begin{equation}\label{eq:M_spectral} 
M(t) = \int_{\mathbb{R}} e^{-i \omega t}\mu(\diff \omega),
\end{equation}
where $\mu(.)$ is a finite symmetric Borel measure on the real line with the Borel sets %${\cal B}(\mathbb{R})$ 
as measurable sets. In particular such equilibrium correlation functions are real and fulfill $|M(t)| \leq M(0) = \mu(\mathbb{R}) =: M$. 

The equation of motion, Eq.~\eqref{eq:harmonic_oscillator}, naturally emerges in statistical physics for dynamic phenomena either as a phenomenological model, or as derived by the Zwanzig-Mori projection operator technique~\cite{KuboBuch:1991,Hansen:Theory_of_Simple_Liquids,Goetze:Complex_Dynamics}, or field-theoretic methods. Despite its formal similarity to an harmonic oscillator equation, the description is not restricted to harmonic interactions. Rather the spirit is that the complex many-particle interactions are 
included in $M(t)$, referred to as memory kernel, self-energy, or mass operator in different physics communities.    

One  derives the short-time expansion of Eq.~\eqref{eq:harmonic_oscillator} for $t\to 0$
\begin{align}\label{eq:short_time}
 S(t) =& S - J t^2/2 + J \nu J |t|^3/6 \nonumber \\ 
&+ J(M+S^{-1} - \nu J \nu) J t^4/24 +  o(t^4),
\end{align}
where  $o(\cdot)$ denotes the little-$o$ Landau symbol. To path the way for generalizations, we shall display equations such that they are valid also in the case of matrix-valued correlation functions, i.e. if the quantities $S, J, \nu$, etc., do not necessarily commute.

The goal of these notes is  to demonstrate the existence of the long-time limit $\lim_{t\to\infty}S(t)$.

\subsection{Representation of solutions}
First we recall that  $S(t)$  is uniquely determined and corresponds to an equilibrium correlation function thereby introducing some notation and relations that will turn out useful to discuss the long-time limit. The proof is performed most conveniently in the Fourier-Laplace domain and is  basically the easy direction of the representation theorems by Nevanlinna, Herglotz, etc.~\cite{Akhiezer:Classical_Moment_Problem}.  
  
The equations of motion, Eq.~\eqref{eq:harmonic_oscillator}, are readily transformed to 
\begin{align}
\hat{S}(z) =& -[ z S^{-1}+ S^{-1}\hat{K}(z)S^{-1} ]^{-1}, \label{eq:first_eom} 
\end{align}
with 
\begin{align}
\hat{K}(z) =& - [ z J^{-1} + i \nu + \hat{M}(z) ]^{-1}. \label{eq:second_eom}
\end{align}
Since $\hat{M}(z)$ is the Borel-Stieltjes transform of a finite measure, the asymptotic expansion $\hat{M}(z) = - M/z + o(z^{-1})$ holds uniformly in the sector $\delta\leq \arg z \leq \pi-\delta$ for any fixed angle  $\delta >0$ ($<\pi/2)$. All asymptotic expansions in the complex frequencies in the following are to be understood in this sense. Then one finds the corresponding asymptotic expansions
\begin{align}
 \hat{K}(z) =& - J z^{-1} + i J \nu J z^{-2} + J (  \nu J \nu -M) J z^{-3} \nonumber \\
& + o(z^{-3}), \label{eq:high_frequency_K}  \\
 \hat{S}(z) =& - S z^{-1} - J z^ {-3} + i J \nu J z^{-4} \nonumber \\
& + J (\nu J \nu-M - S^{-1}) J z^{-5} + o(z^{-5}). \label{eq:high_frequency}
\end{align}
Clearly, the expansions are equivalent to  the short-time expansion of Eq.~\eqref{eq:short_time}. 

Assuming that $M(t)$ is an equilibrium auto-correlation functions is equivalent to 
$\hat{M}(z)$ fulfills properties (1)-(4). 
Then for $z\in \mathbb{C}_+$ the imaginary part of the denominator in Eq.~\eqref{eq:second_eom} is positive and the inverse is well defined. Thus properties (1) and (2) follow immediately for $\hat{K}(z)$. 
The spectral representation shows that $\hat{M}(z= i \eta) = {\cal O}(\eta^{-1})$ as $\eta\to \infty$, and  one infers $\hat{K}(z=i \eta)= i J/\eta + \mathcal{O}(\eta^{-3})$ as $\eta \to \infty$, which includes property (3). For the last property, write
\begin{align}
\Imag[\hat{K}(z)] = \hat{K}^*(z) \left\{ \Imag[z]J^{-1} + \nu + \Imag[\hat{M}(z)] \right\} \hat{K}(z) >0 ,
\end{align}
for $z\in \mathbb{C}_+$, 
which shows that also property (4) is fulfilled. Hence $\hat{{K}}(z)$ is again the Fourier-Laplace transform of an equilibrium  correlation function. 

Repeating the argument for Eq.~\eqref{eq:first_eom} shows that $\hat{S}(z)$ corresponds again to an equilibrium correlation function. Hence by uniqueness of the Laplace transform, 
the equations of motion yield correlation functions $S(t)$ provided the  kernel $M(t)$ is itself a correlation function. By the spectral representation theorem,  there is thus a finite symmetric Borel measure $\sigma(\cdot)$ such that 
\begin{equation}
S(t) = \int_{\mathbb{R}} e^{-i \omega t} \sigma(\diff \omega),
\end{equation}
in particular $S(0) = S = \sigma(\mathbb{R})$.

\subsection{The spectral measure}
Generally a spectral measure can be uniquely decomposed
\begin{equation}
\sigma = \sigma_\text{ac} + \sigma_\text{s},
\end{equation}
such that $\sigma_\text{ac}$ is absolutely continuous with respect to the Lebesgue measure (i.e. $\sigma_\text{ac}(N)=0$ for any Borel set $N$ of Lebesgue measure zero), and $\sigma_\text{s}$ is singular with respect to the Lebesgue measure (i.e. there is a set $N$ of Lebesgue measure zero with $\sigma_\text{s}(\mathbb{R}\setminus N) =0$) \cite{Teschl:Schroedinger_operators}.  

The goal of this subsection is to show that the singular measure $\sigma_\text{s}$ corresponding to $S(t)$ consists at most of an atom at zero-frequency. The proof relies on the characterization of the minimal support of the singular measure as inferred from the boundary values of $\Imag[\hat{S}(z)]$ towards the real line, see Theorem 3.23 in \cite{Teschl:Schroedinger_operators},  in symbols
\begin{equation}
\text{supp } \sigma_\text{s} = \{ \omega \in \mathbb{R} | \limsup_{\eta \downarrow 0} \Imag[\hat{S}(\omega+ i \eta) ] = \infty \}.
\end{equation}

The first step is to rewrite the equation of motion in the Laplace domain Eqs.~\eqref{eq:first_eom},\eqref{eq:second_eom} as
\begin{equation}
\hat{S}(z) = - S/z + \left[z S^{-1} - z^3 J^{-1} - i z^2 \nu - z^2 \hat{M}(z) \right]^{-1}.
\end{equation}
Next, a partial fraction decomposition shows that 
\begin{equation}\label{eq:estimate}
\Imag[z^2 \hat{M}(z)] + \Imag[z] M  =  \Imag[z] \int \frac{\lambda^2 }{|\lambda- z|^2}  \mu(\diff \lambda) \geq 0,
\end{equation} 
for $z\in\mathbb{C}_+$, in particular $\Imag[z^2 \hat{M}(z)]$ is non-negative close to the real line except for corrections of order ${\cal O}(\Imag[z])$. 

For later purpose we also take the limit $z=\omega+i\eta$ to the real axis and infer the stronger estimate 
\begin{align}\label{eq:Radon} 
\lefteqn{\lim_{\eta\downarrow 0} \Imag[(\omega+i\eta)^2 \hat{M}(\omega+ i\eta) ] \geq \nonumber} \\  
& \geq 
\omega^2 
\lim_{\eta\downarrow 0} \Imag[\hat{M}(\omega+i\eta)] = \frac{1}{\pi}\omega^2 D\mu(\omega) \geq 0.
\end{align}
where $D\mu(\omega)$ denotes the Radon derivative of the Borel measure $\mu(\cdot)$~\cite{Teschl:Schroedinger_operators}.
 
Since $-\Imag[w^{-1}] \leq (\Imag[w])^{-1}$ for any $w\in\mathbb{C}$, the estimate 
\begin{align}\label{eq:master_estimate}
\lefteqn{\Imag[\hat{S}(\omega+ i\eta)] \leq \nonumber } \\
&  \eta S/\omega^2 +  [ \omega^2 \nu +  \Imag[ (\omega+i \eta)^2\hat{M}(\omega+i \eta)]+ 
{\cal O}(\eta)]^{-1},
\end{align}
shows that $\limsup_{\eta\downarrow 0} \Imag[\hat{S}(\omega+i\eta)]$ remains finite for $\omega\neq 0$. Thus $\text{supp } \sigma_\text{s} \subset \{0 \}$. The decomposition into a singular continuous part and a pure point spectrum $\sigma_\text{s} = \sigma_\text{sc} + \sigma_\text{pp}$ allows to exclude the singular continuous part $\sigma_\text{sc} = 0$ and leaves at most a pure point contribution at zero frequency $\sigma_\text{pp} = F \delta_0, F\geq 0$, where $\delta_0$ is the Dirac measure concentrated at the origin.   

The correlation function $S(t)$ thus can be uniquely represented by
\begin{equation}
S(t) = F + \int_{\mathbb{R}} e^{-i \omega t}\sigma_{\text{ac}}(\diff \omega). 
\end{equation}
By the Riemann-Lebesgue lemma, the absolutely continuous part does not contribute at large times, such that the long-time limit is guaranteed to exist and equals
\begin{equation}
\lim_{t\to \infty} S(t) = F = \sigma(\{ 0 \}).
\end{equation}
Quite generally the pure point part at zero frequency can be also obtained from the Fourier-Laplace transform (see G\"otze~\cite{Goetze:Complex_Dynamics}, p. 588, Appendix A, Eq. (A.36)), 
\begin{equation}
\lim_{t\to\infty} S(t) = \lim_{\eta \downarrow 0} \eta \int_0^\infty e^{-\eta t} S(t) \diff t = \lim_{\eta \downarrow 0} (-i\eta) \hat{S}(z=i\eta).
\end{equation}

One can show that for arbitrary  correlation functions $M(t)$  the long-time average defined by
\begin{align}
 \limav_{t\to\infty} M(t) &= \lim_{T\to \infty} \frac{1}{T} \int_0^T M(t) \diff t, 
\end{align}
always exists~\cite{Feller:Probability_2,Goetze:Complex_Dynamics}. Clearly it coincides with the long-time limit if existent. 
Equivalent representation follow from the spectral representation theorem or from the low-frequency behavior of the Laplace transform
\begin{align}
\limav_{t\to \infty} M(t) & = \lim_{\eta \downarrow 0} (-i\eta) \hat{M}(z=i\eta), \\
\limav_{t\to\infty} M(t) &= \mu(\{ 0 \}),
\end{align}
 where $\mu(\cdot)$ again represents the spectral measure associated with $M(t)$. 
Performing the low-frequency limit of the equations of motion in the Laplace domain, Eqs.~(\ref{eq:first_eom},\ref{eq:second_eom}) reveals that the long-time limit of the correlation function $S(t)$ can be evaluated from the long-time average of the memory kernel
\begin{equation}\label{eq:long_time_limits}
 \lim_{t\to\infty } S(t) = S - [S^{-1} + \limav_{t\to\infty} M(t)]^{-1}.
\end{equation}
Hence, the long-time limit can be readily calculated from the long-time average of the memory kernel without solving for the entire time-dependence.

\section{Implications for mode-coupling theory}\label{Sec:MCT}
The mode-coupling theory of the glass transition as developed by G\"otze and collaborators~\cite{Goetze:Complex_Dynamics} deals with a non-linear integro-differential equation, which encodes a plethora of 
non-trivial feature in striking agreement with experiments on the evolution of glassy dynamics. Here we review only a simplified version known as schematic models. In the Newtonian case the starting point is an equation of motion as in Eq.~\eqref{eq:harmonic_oscillator}, yet the memory kernel is given as a non-linear functional of the correlation function itself
\begin{align}
& J^{-1} \ddot{S}(t) + \nu  \dot{S}(t) +  S^{-1} S(t) + \int_0^t  M(t-t')  \dot{S}(t') \diff t' =0 , \nonumber \\
& M(t) = {\cal F}[S(t)] = \sum_{n\in \mathbb{N}} v_n S(t)^n.
\end{align}
The functional ${\cal F}[F] = \sum_{n} v_n F^n$ is assumed to be absolutely monotone, i.e. $v_n\geq 0$ and $\sum_n v_n S^n < \infty$. Again, the equations are subject to the 
initial conditions $S(t=0) = S>0, \dot{S}(t=0)=0$.

The existence and uniqueness of solutions has been demonstrated by Haussmann~\cite{Haussmann:1990}, moreover he proved that the solutions correspond to correlation functions in the sense of Sec.~\ref{Sec:correlation_functions}. The key observation is that the mode-coupling functional ${\cal F}[\cdot]$ maps correlation functions on correlation functions. Assuming that the long-time limit exists $F= \lim_{t\to\infty} S(t)$ it has been shown that the equations display a  covariance property which allows to demonstrate a certain maximum principle~\cite{Houches:1991}. As a consequence the long-time limit can be determined by solving merely an algebraic equation rather than by determining the entire time-dependent solution. Therefore the non-equilibrium state diagram, distinguishing between ergodic 'liquid' states and non-ergodic 'glass' states, can be constructed by a convergent iteration procedure~\cite{Goetze:Complex_Dynamics,Goetze:1995}.  
  
A consequence of the proof in Sec.~\ref{Sec:Model_equations} is that the existence of the long-time limit of the solutions of the mode-coupling theory is guaranteed for $\nu>0$. By Haussmann's proof~\cite{Haussmann:1990} both $S(t)$ and $M(t)$ correspond to a correlation function. Taking therefore Haussmann's $M(t)$ as known, the proof of the previous section shows that $F = \lim_{t\to\infty } S(t)$ 
indeed exists.   

The case of pure relaxational dynamics is obtained essentially by dropping the second time derivative in the equation of motion and specifying only the initial value of the correlation function, -- the initial value of the derivative then follows from the equation of motion. Then it has been shown that the solutions are correlation functions that are also completely monotone~\cite{Goetze:1995,Franosch:2002,Goetze:Complex_Dynamics}, i.e. by Bernstein's theorem~\cite{Feller:Probability_2} they can be represented as a Lebesgue-Stieltjes integral of relaxing exponentials. Then the existence of the long-time limit is guaranteed, in addition to the conclusions inferred for the non-equilibrium state diagram  also the long-time behavior has been characterized.

\section{Conclusion}
We have shown that the equation of motion displays unique solutions that correspond to equilibrium correlation functions, provided the memory kernel displays the same properties. 
Furthermore  the long-time limit of these solutions always exists and can be related to the long-time average of the memory kernel. The key ingredient is to rule out a  singular continuous spectrum 
in the spectrum associated with the correlation function $S(t)$, which has been achieved by relying on a characterization in terms of the boundary values of the imaginary part of the Laplace transform. 

In the context  of liquid dynamics the proof corroborates that the mode-coupling theory encompassing the phenomenology of the evolution of structural relaxation is built on mathematically solid grounds. More precisely, we have proven the assumed existence of the long-time limit for Newtonian dynamics which is key to characterize idealized glass states and the associated non-equilibrium state diagram.

The generalizations to matrix-valued quantities as they naturally occur considering several fluctuating variables, respectively their covariance matrix, 
is straightforward and the formulation chosen suggests that all equations can also be interpreted for matrices. Then the concept of non-negativity has to be adapted to positive semi-definite for matrices, etc., see e.g. Ref.~\cite{Franosch:2002,Lang:2013}. Matrix-valued correlation functions occur for example considering mixtures of different species~\cite{Hansen:Theory_of_Simple_Liquids,Goetze:Complex_Dynamics,Franosch:2002}.
 Several relaxation channels  emerge naturally for molecular fluids~\cite{Franosch:1997,Scheidsteger:1997} or in confined geometry~\cite{Lang:2010,Lang:2012,Lang:2013} and  yield
 a different mathematical structure, nevertheless it appears feasible to demonstrate the existence of the long-time limit  also for this case.

The key observation was that the singular spectrum could be ruled out, since in $\nu +  \Imag[ \hat{M}(z)] > 0$ the  term $\nu>0$ dominates for sufficiently small 
$\Imag[z]>0$. Conventionally this $\nu$-term is considered to describe processes faster than the time scales of interest, and one may argue that it constitutes an idealization and should not be present in the  microscopic problem. Then one should put $\nu$ to zero and consider these fast processes to be included in $M(t)$, resp. $\hat{M}(z)$. Then Eqs.~\eqref{eq:Radon},\eqref{eq:master_estimate} shows that  the conclusions still hold 
 if $\lim_{\eta\downarrow 0} \Imag[ \hat{M}(z=\omega+ i\eta)]>0$,  i.e. if there are no spectral gaps associated with $M(t)$. Then the correlation function $S(t)$ still displays a long-time limit, which can be obtained from Eq.~\eqref{eq:long_time_limits}.  
 
One may wonder what kind of correlation functions can be obtained in general from an equation of motion of the type of a generalized harmonic oscillator, or conversely, which correlation functions can be represented as solutions of generalized harmonic oscillator equations. Clearly, a necessary condition is that the short-time expansion, Eq.~\eqref{eq:short_time}, respectively, the high-frequency expansion, Eq.~\eqref{eq:high_frequency}, holds with parameters $J>0, \nu>0, M>0$. Then, it is shown in Appendix~\ref{Sec:Appendix} that the representations Eqs.~(\ref{eq:first_eom},\ref{eq:second_eom}) in the Fourier-Laplace domain hold,  such that $\hat{M}(z)$ satisfies conditions (1)-(3) and $\nu+ \Imag[\hat{M}(z)]>0$ for $z\in \mathbb{C}_+$. Going through the proof again reveals that the  long-time limit $S(t)$ still exists provided for each fixed $\Real[z]$, the expression 
$\Imag[ i\nu z^2+  z^2 \hat{M}(z)]>0$ remains separated from zero for $\Imag[z] \downarrow 0$.  Then the value of the long-limit   can be inferred from an equation analogous to Eq.~\eqref{eq:long_time_limits}, where the long-time average is replaced by $\lim_{\eta\downarrow 0} (-i\eta) \hat{M}(z=i\eta)$, the existence of the limit being guaranteed from the one of $\lim_{\eta\downarrow 0} (-i\eta) \hat{S}(z=i\eta)$. In the case that already $\Imag[\hat{M}(z)] \geq 0$ for $z \in \mathbb{C}_+$,  the kernel $\hat{M}(z)$ corresponds itself to a correlation function $M(t)$, and $S(t)$ also fulfills the generalized harmonic oscillator equation in the temporal domain, Eq.~\eqref{eq:harmonic_oscillator}.

Last, one may consider also correlation functions for stationary stochastic processes that do not correspond to equilibrium states. These correlation functions are then in general complex-valued but are still Fourier transforms of a measure, albeit not necessarily a symmetric one~\cite{Feller:Probability_2}. The corresponding Laplace transforms still fulfill properties (1),(3),(4), yet property  (2) may be violated. In particular the imaginary part $\Imag[\hat{S}](z)$ will not be a symmetric function for frequencies approaching the real line, which reflects that absorption and emission of quanta in a scattering process is no longer equally likely~\cite{Hansen:Theory_of_Simple_Liquids}. Going through the proofs again, one verifies that if we relax the assumption for the kernel $M(t)$ to be a correlation function in a non-equilibrium steady state, then so will be the solutions  $S(t)$ of the equations of motion, Eq.~\eqref{eq:harmonic_oscillator}, and again the long-time limit $\lim_{t\to\infty} S(t)$ is guaranteed to exist.

\begin{acknowledgments}
I am indebted  to  Wolfgang G\"otze and Rolf Schilling for priceless discussions on the mathematical properties of the mode-coupling theory of the glass transition. 
This work has been supported by the
Deutsche Forschungsgemeinschaft DFG via the  Research Unit FOR1394 ``Nonlinear Response to
Probe Vitrification''.
\end{acknowledgments}

%\end{document}

\appendix
\section{Representation lemma}\label{Sec:Appendix}

In this Appendix
we adapt a representation lemma (see Akhiezer~\cite{Akhiezer:Classical_Moment_Problem}, p.111, Lemma 3.3.6) for a correlation function $S(t)$ displaying a short-time expansion of Eq.~\eqref{eq:short_time}, respectively, the high-frequency expansion, Eq.~\eqref{eq:high_frequency},  with parameters $J>0, \nu>0, M>0$. The adaptation permits to include the term due to $\nu>0$, in the representation. 

 Since $S(t) = \int_\mathbb{R} \exp(-i \omega t) \sigma(\diff \omega)$ is assumed to be a correlation function, $\hat{S}(z)$ fullfills conditions (1)-(4), hence in particular, they are Nevanlinna functions~\cite{Akhiezer:Classical_Moment_Problem}. Then by the Hamburger-Nevanlinna theorem (see Akhiezer~\cite{Akhiezer:Classical_Moment_Problem}, p.95, Theorem 3.2.1) this implies that $J = \int_\mathbb{R} \omega^2 \sigma(\diff \omega)$, which also entails the existence $\int_\mathbb{R} \omega \sigma(\diff \omega)=0$, while $\int_\mathbb{R} \omega^4 \sigma(\diff \omega) = \infty$ as signaled by the imaginary prefactor of the ${\cal O}(z^{-4})$ term. The high-frequency expansion for the Nevanlinna function $\hat{S}(z)$ implies a representation of the form of Eq.~\eqref{eq:first_eom}
with a Nevanlinna function $\hat{K}(z)$ (see Akhiezer~\cite{Akhiezer:Classical_Moment_Problem}, p.111, Lemma 3.3.6), i.e. it reflects properties (1),(4). Property (2) is inherited directly from the corresponding property for $\hat{S}(z)$ by the representation. 
Expanding in powers of $z^{-1}$ reveals the high-frequency expansion Eq.~\eqref{eq:high_frequency_K}, in particular, demonstrating property (3). Thus $\hat{K}(z)$ corresponds again to a correlation function. 

The presence of the term $i  J \nu J z^{-2}$ in Eq.~\eqref{eq:high_frequency_K} spoils the direct application of  the lemma  again to find the representation of Eq.~\eqref{eq:second_eom} with a correlation function $\hat{M}(z)$, but a slight modification of the proof is sufficient for our purposes.

\begin{lemma}
A function $\hat{K}(z)$ corresponding to a correlation function and high-frequency series
\begin{align}
 \hat{K}(z) =& - J z^{-1} + i J \nu J z^{-2} + J (  \nu J \nu -M) J z^{-3} \nonumber \\
& + o(z^{-3}), \label{eq:lemma_assumption} 
\end{align}
with parameters $J>0,\nu>0, M>0$
 can be represented for $z\in \mathbb{C}_+$ as
\begin{equation}
\hat{K}(z) = - [ z J^{-1} + i \nu + \hat{M}(z) ]^{-1}, \label{eq:lemma}
\end{equation}
where $\hat{M}(z)$ fulfills properties (1),(2),(3) and $\nu+\Imag[\hat{M}(z)] \geq 0$ for $z\in \mathbb{C}_+$. 
\end{lemma}

\emph{Proof:} Since $\hat{K}(z)$ corresponds to a correlation function, there is  a (symmetric) finite Borel measure $\kappa$ such that the representation
\begin{equation}
 \hat{K}(z) = \int_\mathbb{R} \frac{\kappa(\diff \omega)}{\omega- z},
\end{equation}
holds, in particular from Eq.~\eqref{eq:lemma_assumption}  one infers $J = \int_\mathbb{R} \kappa(\diff \omega)$. Furthermore, its imaginary part
\begin{equation}
 \Imag{[\hat{K}(z)]} = \int_\mathbb{R} \frac{\Imag[z] }{|\omega- z|^2} \kappa(\diff \omega),
\end{equation}
is harmonic and non-negative for $z\in\mathbb{C}_+$. By the mean-value property of harmonic functions a zero can occur only if $\hat{K}(z) \equiv 0$ for all $z\in \mathbb{C}_+$ which is excluded by the asymptotic expansion, Eq.\eqref{eq:lemma_assumption}. 

Then   the function $\hat{M}(z)$ obtained by inverting the representation
\begin{equation}
 i \nu + \hat{M}(z) = - z J^{-1} - \hat{K}(z)^{-1} ,
\end{equation}
is well defined and properties (1),(2) follow for $\hat{M}(z)$. Expanding in powers of $z^{-1}$ shows that $\hat{M}(z) = - M z^{-1} + o(z^{-1})$ for large frequencies, i.e. property (3) follows. 
Last, taking the imaginary part yields
\begin{equation}
\nu +  \Imag[\hat{M}(z)] = - J^{-1} \Imag[z] + \hat{K}^*(z)^{-1} \Imag[\hat{K}(z)] \hat{K}(z)^{-1}
\end{equation}
The r.h.s. is non-negative for $z\in \mathbb{C}_+$ as can be seen as follows. Define the function 
\begin{equation}
 x(\omega) = (\omega-z)^{-1} \hat{K}(z)^{-1} - J^{-1}.
\end{equation}
Then $\int_\mathbb{R} |x(\omega)|^2 \kappa(\diff\omega)\geq 0$ and upon expanding the square one finds 
\begin{equation}
 \hat{K}^*(z)^{-1} \int_\mathbb{R} \frac{\kappa(\diff  \omega)}{|\omega-z|^2} \hat{K}(z)^{-1}- J^{-1} \geq 0, 
\end{equation}
demonstrating $\nu + \Imag[\hat{M}(z)]\geq 0$ in the upper complex half plane. The asymptotic expansion, Eq.~\eqref{eq:lemma_assumption}, together with the representation, Eq.~\eqref{eq:lemma}, shows that $i\nu +\hat{M}(z)$ does not vanish identically. By the mean-value property of analytic functions this implies even $\nu + \Imag[\hat{M}(z)]>  0$ for $z\in \mathbb{C}_+$.
$\square$

Note that in general it does not follow that  $\hat{M}(z)$ corresponds to a correlation function $M(t)$, a counter example can be constructed involving  for $M(t)$  the difference of two relaxing exponentials.  

The proof is again suitable to generalizations for a matrix-valued quantities, c.f. Ref.~\cite{Lang:2013}.

% %\bibliographystyle{apsrev}
%\bibliographystyle{apsrev4-1}

%\bibliography{mct}

%merlin.mbs apsrev4-1.bst 2010-07-25 4.21a (PWD, AO, DPC) hacked
%Control: key (0)
%Control: author (72) initials jnrlst
%Control: editor formatted (1) identically to author
%Control: production of article title (-1) disabled
%Control: page (0) single
%Control: year (1) truncated
%Control: production of eprint (0) enabled
%

\end{document}